\documentclass[onecolumn, aps, nofootinbib, prd, preprint, floats, floatfix, amsmath, amssymb, superscriptaddress, preprintnumbers]{revtex4-2}
\PassOptionsToPackage{dvipsnames}{xcolor}
\usepackage{array}[=2016-10-06]
\usepackage{slashed}
\usepackage{graphicx}
\usepackage{subcaption}
\usepackage{dcolumn}
\usepackage{bm}
\usepackage{amssymb}
\usepackage{tikz}
\usepackage{tikz-feynman}
\usepackage[utf8]{inputenc}
\usepackage{yfonts,amsmath,amsthm,amsfonts,amssymb,amscd, ulem}
\usepackage{enumerate}
\usepackage{fancyhdr}
\usepackage{mathtools}
\usepackage{amsmath}
\usepackage{autobreak}
\usepackage{mathrsfs}
\usepackage{cancel}
\usepackage{slashed}
\usepackage{bigints}
\usepackage[flushleft]{threeparttable}
\usepackage[colorlinks=true, citecolor=purple, linkcolor=blue]{hyperref}
\usepackage{url}
\usepackage{makecell,booktabs}
\usepackage{braket}
\usepackage{relsize}
\usepackage{multirow}
\usepackage{verbatim}
\usepackage{txfonts}
\usepackage{slashed}
\usepackage{upgreek}
\usepackage{extarrows}
\usepackage{appendix}
\usepackage[T1]{fontenc}
\usepackage{setspace}
\usepackage{subcaption}

\usepackage{xcolor}
\definecolor{c1}{HTML}{1bd1a5}
\usepackage{graphicx}   
\usepackage{caption}    
\usepackage{subcaption} 
\usepackage{subcaption}
\usepackage{cleveref}
\usepackage{orcidlink}
\begin{document}

\title{An Axial $U_A(1)_{L_\mu-L_\tau}$: UV Completion and Experimental Searches}

\author{Rundong Fang}
\email{rundongfang@pku.edu.cn}
\affiliation{School of Physics and State Key Laboratory of Nuclear Physics and Technology, Peking University, Beijing 100871, China}

\author{Jinhui Guo}
\email{guojh23@buaa.edu.cn}
\affiliation{School of Physics, Beihang University, Beijing 100191, China}

\author{Ming Li \orcidlink{0009-0007-7322-010X}}
\email{by2419116@buaa.edu.cn}
\affiliation{School of Physics, Beihang University, Beijing 100191, China}

\author{Jia Liu \orcidlink{0000-0001-7386-0253}}
\email{jialiu@pku.edu.cn}
\affiliation{School of Physics and State Key Laboratory of Nuclear Physics and Technology, Peking University, Beijing 100871, China}
\affiliation{Center for High Energy Physics, Peking University, Beijing 100871, China}

\author{Xiao-Ping Wang \orcidlink{0000-0002-2258-7741} }
\email{hcwangxiaoping@buaa.edu.cn}
\affiliation{School of Physics, Beihang University, Beijing 100191, China}

\author{Yiheng Xiong \orcidlink{0009-0006-1377-1228} }
\email{yihenghht@buaa.edu.cn}
\affiliation{School of Physics, Beihang University, Beijing 100191, China}

\begin{abstract}
We propose an anomaly-free and renormalizable axial $U_A(1)_{L_\mu-L_\tau}$ model and study its experimental signatures for $A'$ masses from the MeV scale to the TeV scale. The opposite charges of the left- and right-handed charged leptons forbid the usual muon and tau Yukawa interactions. Their masses are instead generated by a singlet scalar and heavy vector-like leptons through a universal-seesaw mechanism. We focus on heavy vector-like leptons, small light--heavy mixing, and $m_s\gtrsim10~\mathrm{GeV}$. In this limit, the observables considered here depend mainly on $(m_{A'},g_X)$, while the other model parameters are restricted by mixing and perturbativity. We confront this benchmark with current experimental searches. For neutrino trident production, our finite-$m_\mu$ calculation shows that $A'$ modifies the axial weak coefficient, rather than the vector coefficient relevant to the usual $L_\mu-L_\tau$ model. The longitudinal mode enhances muon bremsstrahlung and gives a negative contribution to $(g-2)_\mu$; the latter dominates over the scalar contribution in our benchmark. Combining these results with invisible meson decays and four-muon resonance searches, we summarize the phenomenological constraints in the $(m_{A'},g_X)$ plane. For $m_{A'}\gg m_\mu$, vector and axial final-state-radiation rates become nearly identical, so the corresponding collider limits can be obtained by rate matching. At a future muon collider, the total rate alone does not fully resolve the interaction structure, whereas angular distributions, especially the forward--backward asymmetry in $\mu^+\mu^-\to\tau^+\tau^-$, retain direct sensitivity to chirality.
\end{abstract}

\makeatletter
\@booleanfalse\titlepage@sw
\@booleanfalse\preprintsty@sw
\makeatother
\maketitle
\makeatletter
\@booleantrue\preprintsty@sw
\makeatother
\tableofcontents

\section{Introduction}
\label{sec:introduction}

Light neutral spin-one bosons provide a minimal and experimentally accessible extension of the Standard Model (SM). Besides the kinetically mixed dark photon~\cite{Holdom:1985ag}, anomaly-free differences of lepton-flavor numbers single out predictive leptophilic gauge symmetries, of which $U(1)_{L_\mu-L_\tau}$ is the best studied~\cite{Foot:1990mn,He:1990pn,He:1991qd}. In its conventional realization, the new boson couples vectorially to charged muons and taus and left-handedly to their neutrinos. This economical structure has been connected to neutrino masses and mixing, a positive contribution to the muon anomalous magnetic moment, dark-sector interactions, and multi-lepton collider signatures~\cite{Baek:2001kca,Ma:2001md,Heeck:2011wj,Harigaya:2013twa,Altmannshofer:2014cfa,Crivellin:2015mga,Altmannshofer:2016brv}. It also defines the standard two-parameter $(m_{Z'},g')$ benchmark against which most searches are reported.

The corresponding experimental program now spans several complementary observables. Neutrino trident production provides a particularly robust low-mass constraint~\cite{Altmannshofer:2014pba}, while general collider studies identified multi-lepton final states as the natural probes of leptophilic vectors~\cite{delAguila:2014soa}. At electron colliders, final-state radiation from a muon gives either a visible four-muon resonance or an invisible recoil-mass signal; these channels were developed in phenomenological studies~\cite{Araki:2017wyg,Brown:2024jah} and pursued by BaBar and Belle II~\cite{BaBar:2016sci,Belle-II:2019qfb,Belle-II:2022yaw,Belle-II:2024wtd}. At the LHC, CMS searched for $Z\to\mu^+\mu^-Z'$ with $Z'\to\mu^+\mu^-$~\cite{CMS:2018yxg}, whereas the NA64 muon program probes invisibly decaying mediators through missing energy and momentum~\cite{NA64:2024klw}; proposed muon missing-momentum experiments extend the same strategy~\cite{Kahn:2018cqs}. These results constitute essential experimental anchors, but their published coupling contours assume the conventional vector current, together with benchmark-dependent widths and branching fractions. They therefore cannot in general be read as model-independent limits on an axial mediator.

The distinction is physical rather than semantic. Replacing the charged-lepton vector current by an axial current changes which SM four-fermion coefficient is shifted in neutrino trident production, reverses the role of $(g-2)_\mu$ by producing a negative and longitudinally enhanced contribution, and can alter production kernels, decay widths, and angular acceptance in visible or missing-energy searches. It also removes some familiar vector-model intuition: in the absence of an additional ultraviolet source, a purely axial charged-lepton current does not generate the ordinary parity-even photon--$A'$ kinetic mixing at one loop. Existing studies of light axial forces, anomaly-free axial $Z'$ theories, nuclear-transition anomalies, and axial muon-force searches have exposed parts of this phenomenology and its consistency conditions~\cite{Kahn:2016vjr,Ismail:2016tod,Kozaczuk:2016nma,Hicyilmaz:2022owb,Batell:2011qq,Davoudiasl:2012ag,Fayet:2020bmb,Baruch:2022esd,Fayet:2024ddk}. Recent flavor-specific chiral and universal-seesaw constructions further emphasize that a genuine axial gauge interaction is inseparable from its symmetry-breaking scalars and heavy-fermion sector~\cite{Prajapati:2026tfv,Dutta:2026dnf}. What remains missing is a unified axial $L_\mu-L_\tau$ treatment in which this UV structure and the experimental reinterpretation are developed together.

We address this gap with an anomaly-free, leptophilic $U_A(1)_{L_\mu-L_\tau}$ construction in which opposite left- and right-handed charged-lepton charges produce an axial current, while the neutrino current remains left-handed. Because the symmetry forbids the diagonal SM Yukawa operators, a singlet scalar and heavy vector-like leptons generate the charged-lepton masses and correlate the low-energy theory with the symmetry-breaking sector. On this UV-consistent footing, we derive the mass-basis gauge and scalar interactions, calculate neutrino trident production with finite $m_\mu$, reinterpret invisible pion-decay and muon-beam missing-energy searches, and clarify the axial $(g-2)_\mu$ bound and its relation to the UV parameter space. We also compare axial and vector signals at a muon collider, distinguishing approximately rescalable rate observables from angular distributions that retain direct sensitivity to chirality.

This paper is organized as follows. In Sec.~\ref{sec:model-setup} we present the UV model and its mass-basis interactions. In Sec.~\ref{sec:phenomenology} we derive and reinterpret the relevant low-energy and intensity-frontier constraints. In Sec.~\ref{sec:muon-collider} we study future muon-collider probes and axial--vector discrimination. We summarize our results in Sec.~\ref{sec:conclusion}.

\section{Model Setup}
\label{sec:model-setup}

\subsection{An anomaly-free axial $U_A(1)_{L_\mu-L_\tau}$ completion}

The conventional $U(1)_{L_\mu-L_\tau}$ extension is anomaly free with the
Standard Model (SM) fermion content and gives vector couplings to charged
leptons.  An axial interaction instead requires opposite charges for the
left- and right-handed charged leptons.  This assignment forbids the SM
Yukawa operators $\bar L_\mu H\mu_R$ and $\bar L_\tau H\tau_R$; hence a
consistent axial model must explain charged-lepton mass generation as well as
gauge-anomaly cancellation.

We extend the SM gauge group by $
U_A(1)_{L_\mu-L_\tau}$ and introduce two vector-like charged leptons,
$E_\mu$ and $E_\tau$, and a complex SM-singlet scalar $S$.  The scalar vacuum
expectation value (VEV) breaks $U_A(1)_{L_\mu-L_\tau}$, while the heavy leptons generate the
muon and tau masses through a universal-seesaw mechanism.  Our charge
convention is given in Table~\ref{tab:charge-assignment}.  Only the product of
the gauge coupling and the charge normalization is physical; we define
$g_X\equiv g_{A'}Q_\ell$ and later set $Q_\ell=1$.

\begin{table}[htbp]
    \centering
    \begin{tabular}{cccc}
    \hline
    Field & $SU(2)_L$ & $U(1)_Y$ & $U_A(1)_{L_\mu-L_\tau}$ \\
    \hline
    $L_{\mu}=\left(\nu_{\mu}, \mu_L\right)^T$ & $\mathbf{2}$ & $-1 / 2$ & $Q_{\ell}$ \\
    \hline
    $\mu_R$ & $\mathbf{1}$ & -1 & $-Q_\ell$ \\
    \hline
    $E_\mu\left(E_{\mu L}, E_{\mu R}\right)$ & $\mathbf{1}$ & -1 & $Q_\ell$\\
    \hline
    $L_{\tau}=\left(\nu_{\tau}, \tau_L\right)^T$ & $\mathbf{2}$ & $-1 / 2$ & $-Q_\ell$ \\
    \hline
    $\tau_R$ & $\mathbf{1}$ & -1 & $Q_\ell$ \\
    \hline
    $E_\tau\left(E_{\tau L}, E_{\tau R}\right)$ & $\mathbf{1}$ & -1 & -$Q_\ell$\\
    \hline
    $H$ & $\mathbf{2}$ & $1 / 2$ & 0 \\
    $S$ & $\mathbf{1}$ & 0 & $Q_S=2Q_\ell$\\
    \hline
    \end{tabular}
    \caption{Field content and $U_A(1)_{L_\mu-L_\tau}$ charge assignment for the axial $L_\mu-L_\tau$ construction. The electron sector is neutral under $U_A(1)_{L_\mu-L_\tau}$.}
    \label{tab:charge-assignment}
\end{table}

An important feature of this construction is that all gauge and mixed gravitational anomalies cancel exactly, even though each lepton generation is individually chiral under the new gauge symmetry. As an illustration, the cubic anomaly satisfies \cite{Ismail:2016tod}
\begin{equation}
\mathcal{A}_{[U_A(1)_{L_\mu-L_\tau}]^3}
\propto
2(+1)^3
-(-1)^3
+2(-1)^3
-(+1)^3
=0,
\end{equation}
where the factor of two originates from the $SU(2)_L$ doublets and the right-handed fermions contribute with the opposite sign. The mixed gravitational anomaly cancels in the same manner. Consequently, the model provides a renormalizable and anomaly-free realization of an axial $U_A(1)_{L_\mu-L_\tau}$ gauge symmetry.

With the field content fixed by Table~\ref{tab:charge-assignment}, the renormalizable Lagrangian of the UV-complete theory is given by
\begin{equation}\label{eq:Lagrangian}
\begin{aligned}
\mathcal{L}_{UV}\supset& -\frac{1}{4}A'_{\alpha\beta}A'^{\alpha\beta}+\sum_{\alpha=\mu,\tau}\left(i\bar{L}_\alpha \gamma^\nu D_{\nu} L_\alpha + i\bar{\ell}_{\alpha R} \gamma^\nu D_{\nu}\ell_{\alpha R} + i\bar{E}_{\alpha}\gamma^\nu D_{\nu} E_{\alpha}\right)  \\
& -y_\mu \bar L_\mu H E_{\mu R} -\lambda_\mu S \bar E_{\mu L}\ell_{\mu R} -M_{E\mu}\bar E_{\mu L}E_{\mu R} + h.c.\\
&-y_\tau \bar L_\tau H E_{\tau R} -\lambda_\tau S^{*}\bar E_{\tau L}\ell_{\tau R} -M_{E\tau}\bar E_{\tau L}E_{\tau R} + h.c.\\
&+ (D_{\nu}S)^\dagger (D^{\nu}S) - V(S),
\end{aligned}
\end{equation}
where $A'_{\alpha\beta}=\partial_\alpha A'_\beta - \partial_\beta A'_\alpha$ is the field-strength tensor, the covariant derivative is given by $D_\nu = \partial_\nu -ig_Y Y B_\nu  -ig_W T^iW_\nu^i - ig_{A'} Q_{A'} A'_\nu$, where $B_\nu$, $W_\nu^i$ and $A'_\nu$ represent the gauge fields for the ${U(1)}_Y$, ${SU(2)}_L$ and ${U_A(1)}_{L_\mu-L_\tau}$ groups, respectively. The constants $g_Y$, $g_W$ and $g_{A'}$ are their associated coupling constants. And, $V(S)$ is the self-interacting potential of $S$. Furthermore, the ordinary diagonal SM Yukawa operator $\bar L_\ell H\ell_R$ is forbidden due to the charge assignment. Consequently, the interactions between charged leptons and Higgs in this model necessarily arise via the new fields, $E$ and $S$. Integrating out the heavy field $E$ at the matching scale $\Lambda=M_E$ generates the following dimension-5 effective operators:
\begin{equation}\label{eq:5-d}
    \mathcal{L}_{\text{5-d}} =
    -\frac{c_\mu}{\Lambda} S \bar{L}_\mu H\mu_R
    -\frac{c_\tau}{\Lambda} S^* \bar{L}_\tau H\tau_R
    +h.c.,
\end{equation}
where, at the tree level, $\frac{c_\alpha}{\Lambda}\simeq \frac{y_\alpha \lambda_\alpha}{M_{E\alpha }}$ for $\alpha=\mu,\tau$.

After the electroweak and $U_A(1)_{L_\mu-L_\tau}$ spontaneous symmetry breaking, the scalar fields are given by $H=\frac{1}{\sqrt{2}}\left (\begin{array}{c} 0 \\ v_h+h \end{array}\right)$ and $S=\frac{1}{\sqrt{2}} (v_s + s)$, where $v_{h/s}$ denotes the VEVs of the Higgs and scalar $S$, respectively. The kinetic term
$(D_\nu S)^\dagger(D^\nu S)$ gives the new gauge-boson mass as following
\begin{align}
m_{A'}=|Q_S|g_{A'}v_s=2|Q_\ell|g_{A'}v_s\equiv2|g_X|v_s,
\end{align} 
with $g_X\equiv g_{A'}Q_\ell$. And the mass terms for leptons can be derived as
\begin{equation}
\mathcal{L}_{\mathcal M}=-
\begin{pmatrix}
	\overline{\ell^{\mu}_L},  &
	\overline{E^\mu_L}, &
	\overline{\ell^\tau_L}, &
	\overline{E^\tau_L}
\end{pmatrix}
\begin{pmatrix}
	0 & y_{\mu}v_h/\sqrt2  & 0 & 0 \\
	\lambda_\mu v_s/\sqrt2 & M_{E\mu} & 0 & 0\\
	0 & 0 & 0 & y_{\tau}v_h/\sqrt2\\
    0 & 0 &  \lambda_\tau v_s/\sqrt2 & M_{E\tau}
\end{pmatrix}
\begin{pmatrix}
	\ell^{\mu}_R \\
	E^\mu_R \\
	  \ell^\tau_R \\
	  E^\tau_R \\
\end{pmatrix}
+h.c. = -\bar{\chi}_L \mathcal{M} \chi_R + h.c.
\label{eq:4by4-mass-no-mutau}
\end{equation}
Here, $\chi_L=(\ell^{\mu}_L,E^\mu_L,\ell^\tau_L, E^\tau_L)^T$, $\chi_R=(\ell^{\mu}_R,E^\mu_R,\ell^\tau_R, E^\tau_R)^T$. Because the
$4\times4$ matrix $\mathcal M$ contains independent muon and tau blocks, each
flavor can be diagonalized separately.  It is sufficient to display the muon
block explicitly; the tau block follows by the corresponding change of flavor
labels.  The muon mass term is
\begin{equation}
\mathcal{L}_{\mathcal{M},\mu}
=-\overline{\chi^\mu_L}\,\mathcal{M}_\mu\,\chi^\mu_R+h.c.
=-\overline{\chi^\mu_L}
\begin{pmatrix}
0 & m_L\\
m_R & M_{E\mu}
\end{pmatrix} \chi^\mu_R+h.c.,
\label{eq:mu-2by2-mass}
\end{equation}
with $\chi^\mu_L=(\ell^\mu_L, E^\mu_L)^T$, $\chi^\mu_R =
(\ell^\mu_R,E^\mu_R)^T$, and $m_L=\frac{y_\mu v_h}{\sqrt2},
m_R=\frac{\lambda_\mu v_s}{\sqrt2}$. Assuming these mass entries are real, the mass matrix can be bi-orthogonally diagonalized as
\begin{equation}
U_{\mu L}^T\mathcal{M}_\mu U_{\mu R} =
\begin{pmatrix}
m_\mu &0\\
0&m_{E\mu}
\end{pmatrix},~ {\rm with}~
\chi^\mu_L=U_{\mu L} \chi^{\mu \prime}_L~{\rm and}~
\chi^\mu_R=U_{\mu R} \chi^{\mu \prime}_R,
\end{equation}
with $U_{\mu L/\mu R}=\begin{pmatrix}\cos\theta_{\mu L/\mu R} &\sin\theta_{\mu L/\mu R}\\-\sin\theta_{\mu L/\mu R} & \cos\theta_{\mu L/\mu R} \end{pmatrix}$. The corresponding mass eigenvalues are
\begin{equation}
m_{\mu/E\mu}^2=
\frac{1}{2}
\left[
m_L^2+m_R^2+M_{E\mu}^2 \mp
\sqrt{(m_L^2+m_R^2+M_{E\mu}^2)^2-4m_L^2m_R^2}
\right].
\label{eq:exact-masses-2by2}
\end{equation}
The mixing angles can be written as
\begin{equation}
\tan2\theta_{\mu L}= \frac{2m_L M_{E\mu}}{M_{E\mu}^2+m_R^2-m_L^2},
\qquad
\tan2\theta_{\mu R}=\frac{2m_R M_{E\mu}}{m_L^2+M_{E\mu}^2-m_R^2}.
\label{eq:mixing-angles-2by2}
\end{equation}
In the limit $M_{E\mu}\gg m_L, m_R$, to the leading order, the muon block masses and mixing angles can be simplified to
\begin{equation}
\begin{aligned}
    &m_\mu^2\simeq\frac{m_L^2 m_R^2}{M_{E\mu}^2},~ m_{E\mu}^2\simeq M_{E\mu}^2+(m_L^2 + m_R^2)\\
    &\theta_{\mu L}\simeq \frac{m_L}{M_{E\mu}},~ \theta_{\mu R}\simeq \frac{m_R}{M_{E\mu}},
\end{aligned}
\end{equation}
The first relation shows explicitly how the light muon mass is suppressed by
the heavy vector-like mass.  It also agrees with the mass obtained from the
dimension-5 operator in Eq.~\eqref{eq:5-d} after symmetry breaking.  The last
two relations show that the same hierarchy suppresses the mixing of the light
and heavy states.

\subsection{Interactions in the mass basis}

The mass rotations affect both the gauge and scalar interactions. It is therefore useful to organize the interaction Lagrangian. We can decompose the interaction into two parts: gauge interaction and scalar interaction,
\begin{equation}
    \mathcal{L}_{\rm int}\supset \mathcal{L}_{\rm gauge} + \mathcal{L}_{\rm scalar}.
\end{equation}
The relevant gauge Lagrangian part can be expanded as
\begin{equation}
\begin{aligned}
\mathcal{L}_{\rm gauge}
&\supset~
\overline{\chi^{\mu\prime}_L}\gamma^\alpha
\bigg[-eA_\alpha\,\mathbf{1}+g_X A'_\alpha\,\mathbf{1} +Z_\alpha
\begin{pmatrix}
g_{\ell_L}^Z c_L^2+g_E^Z s_L^2 & (g_{\ell_L}^Z-g_E^Z)c_Ls_L\\
(g_{\ell_L}^Z-g_E^Z)c_Ls_L & g_{\ell_L}^Z s_L^2+g_E^Z c_L^2
\end{pmatrix}
\bigg] \chi^{\mu\prime}_L\\
&+~\overline{\chi^{\mu\prime}_R}\gamma^\alpha
\bigg[ -eA_\alpha\,\mathbf{1} +g_E^ZZ_\alpha\,\mathbf{1} -g_XA'_\alpha
\begin{pmatrix}
\cos2\theta_{\mu R} & \sin2\theta_{\mu R}\\
\sin2\theta_{\mu R} & -\cos2\theta_{\mu R}
\end{pmatrix}\bigg]\chi^{\mu\prime}_R\\
&+\frac{g_W}{\sqrt2}
\left[
\bar\nu_{\mu L}\gamma^\alpha
\left(c_L\mu'_L+s_L E'_{\mu L}\right)W^+_\alpha
+h.c.
\right]\\
&+ \frac{m_{A'}^2}{v_s}s A'_\alpha A'^\alpha + \frac{m_{A'}^2}{2 v_s^2}s^2 A'^2.
\end{aligned}
\label{eq:mass-basis-NC}
\end{equation} 
where $g_Z=\sqrt{g_W^2+g_Y^2}$, $s_W=g_Y/g_Z$, $c_L=\cos\theta_L$ and $s_L=\sin\theta_L$. The neutral-current couplings of the doublet and singlet charged leptons to the SM $Z$ boson are $g_{\ell_L}^Z=g_Z\left(-\frac12+s_W^2\right), ~g_E^Z=g_Z s_W^2$. We can see that the electromagnetic coupling is unchanged because it is proportional to the identity. The left $A'$ coupling is also unchanged because $\ell_L^\mu$ and $E_L^\mu$ carry the same $U_A(1)_{L_\mu-L_\tau}$ charge. The right $A'$ coupling is not exactly diagonal because $\ell_R^\mu$ and $E_R^\mu$ have opposite $U_A(1)_{L_\mu-L_\tau}$ charges, but the off-diagonal coupling is suppressed by $\sin2\theta_{\mu R}\simeq 2m_R/M_{E\mu}$ in the large $M_{E\mu}$ expansion.

The scalar interactions involving the Higgs boson and $s$ become
\begin{equation}
\begin{aligned}
\mathcal{L}_{\rm scalar} &
\supset-\overline{\chi_L^\mu}
\left[
\frac{m_L}{v_h}\,h
\begin{pmatrix}
0&1\\
0&0
\end{pmatrix}
+
\frac{m_R}{v_s}\,s
\begin{pmatrix}
0&0\\
1&0
\end{pmatrix}
\right]\chi_R^\mu
+h.c. .\\
&=-\overline{\chi'_L}
\left[\frac{m_L}{v_h} h \begin{pmatrix}
c_Ls_R&c_Lc_R\\
s_Ls_R&s_Lc_R
\end{pmatrix}+ \frac{m_R}{v_s} s \begin{pmatrix}
s_Lc_R&-s_Ls_R\\
-c_Lc_R&c_Ls_R
\end{pmatrix}\right]\chi'_R+h.c.,
\label{eq:gauge-basis-scalar-yukawa}
\end{aligned}
\end{equation}
For the light charged lepton, the diagonal couplings are therefore
\begin{equation}
y_{h\mu\mu}=
\frac{m_L}{v_h}c_Ls_R,
\qquad
y_{s\mu\mu}=\frac{m_R}{v_s}s_Lc_R .
\label{eq:light-scalar-yukawa}
\end{equation}
In the seesaw-like limit $M_E\gg m_L,m_R$, these reduce to
\begin{equation}
y_{h\mu\mu}\simeq
\frac{m_\mu}{v_h}
\qquad
y_{s\mu\mu}\simeq \frac{m_\mu}{v_s}.
\label{eq:sm-limit-scalar-yukawa}
\end{equation}
Thus, the SM muon Yukawa coupling is recovered for the observed Higgs in the combined decoupling/alignment limit $M_{E\mu}\gg m_L,m_R$, and with the singlet-like scalar either heavy or weakly mixed, which is consistent with the leading order terms in Eq. \eqref{eq:5-d} after the phase transition. Outside this limit, the new scalar interactions include the singlet coupling proportional to $m_\mu/v_s$, and scalar-mediated transitions between the light lepton and the heavy vector-like lepton.

For the light state ($\mu^\pm$) alone, the diagonal $A'$ coupling is
\begin{equation}
\mathcal{L}_{A'}^{\mu'}
=
A'_\alpha\bar\mu'\gamma^\alpha
\left[
g_X\sin^2\theta_{\mu R}
-g_X\cos^2\theta_{\mu R} \gamma_5
\right]\mu'
+\cdots .
\label{eq:light-muon-Apr}
\end{equation}
Thus, the coupling remains dominantly axial, with only an $\mathcal{O}(\theta_{\mu R}^2)$ vector admixture. The $\tau$ block is obtained by the replacements $\mu\to\tau$ and by reversing the $U_A(1)_{L_\mu-L_\tau}$ charges.

\subsection{Benchmark assumptions and phenomenological parameters}

Before counting the parameters, we specify the flavor structure assumed in our benchmark scenario. Gauge invariance also allows the off-diagonal SM-Higgs Yukawa interaction,
\begin{equation}
    \mathcal{L}\supset
    -y_{\mu\tau}(\bar{L}_{\mu}H \tau_R + \bar{L}_{\tau}H \mu_R)+h.c. .
\label{eq:ymutau-operator}
\end{equation}
which generates non-zero off-diagonal entries
$\mathcal{M}_{13}=\mathcal{M}_{31}=y_{\mu\tau}v_h/\sqrt{2}$
in the lepton mass matrix of Eq.~\eqref{eq:4by4-mass-no-mutau} and induces charged-lepton flavor violation (CLFV) at tree level. This coupling is strongly constrained by searches for
$\tau\to\mu+{\rm Inv}$ \cite{Ibarra:2021xyk},
$\tau\to\mu\mu\mu$ \cite{Ibarra:2021xyk,Hayasaka:2010np,Belle-II:2024sce},
$\tau\to\mu\gamma$ \cite{Belle:2021ysv},
$h\to\mu\tau$ \cite{Harnik:2012pb,CMS:2021rsq}, and
$Z\to\mu\tau$ \cite{ATLAS:2021bdj}.
A systematic overview of CLFV phenomenology in the tau sector is provided in Ref.~\cite{Banerjee:2022xuw}.

From a theoretical perspective, the flavor structure can be motivated by an approximate
$U(1)_{F_\mu}\times U(1)_{F_\tau}$ family symmetry, or an appropriate discrete subgroup, under which the flavor-diagonal Yukawa interactions are allowed while Eq.~\eqref{eq:ymutau-operator} is forbidden. Together with the stringent experimental constraints on CLFV, this motivates our benchmark choice
\begin{equation}
y_{\mu\tau}=0,
\end{equation}
which we adopt throughout this work.

After trading the two parameters of the scalar potential for the symmetry-breaking scale $v_s$ and the physical scalar mass $m_s$, a set of the UV input parameters is
\begin{equation}
    \{v_s, y_\mu, y_\tau, \lambda_\mu,\lambda_\tau, M_{E\mu}, M_{E\tau}, Q_\ell, g_{A'}, m_s \}.
\end{equation}

After electroweak symmetry breaking and after matching to the observed muon and tau pole masses, this model leaves eight independent parameters, which we take as:
\begin{equation}
    \{v_s, \theta_{\mu L}, \theta_{\tau L}, \theta_{\mu R},\theta_{\tau R}, Q_\ell, g_{A'}, m_s \},
    \end{equation}
where the heavy-lepton masses are related to the mixing angles through
\begin{equation}
    m_\mu\simeq \theta_{\mu L} \theta_{\mu R} M_{E\mu}, \qquad
    m_\tau\simeq \theta_{\tau L} \theta_{\tau R} M_{E\tau}.
    \end{equation}

To organize the parameter space, we fix $Q_\ell=1$ and define $g_X\equiv g_{A'}Q_\ell$ throughout the following analysis. We work in the decoupling regime in which the vector-like leptons are much heavier than the off-diagonal entries of the charged-lepton mass matrix, imposing the degenerate benchmark
\begin{equation}
M_{E\mu}=M_{E\tau}\equiv M_E>10~\mathrm{TeV}.
\end{equation}
The degeneracy also controls hypercharge--$U_A(1)_{L_\mu-L_\tau}$ kinetic mixing. The two vector-like leptons have the same hypercharge and opposite $U_A(1)_{L_\mu-L_\tau}$ charges, so their one-loop threshold contributions cancel when $M_{E\mu}=M_{E\tau}$. More generally, a mass splitting would generate a finite contribution proportional to $g_Yg_X\ln(M_{E\tau}/M_{E\mu})$~\cite{Holdom:1985ag}. We take the kinetic-mixing coefficient to vanish at the matching scale; below the heavy-lepton threshold, the pure-axial light-lepton charge assignment does not regenerate it at one loop, up to the mixing-suppressed corrections neglected in our benchmark.
The observed charged-lepton masses then fix the products
\begin{equation}
\theta_{\mu L}\theta_{\mu R}\simeq\frac{m_\mu}{M_E},
\qquad
\theta_{\tau L}\theta_{\tau R}\simeq\frac{m_\tau}{M_E}.
\end{equation}
These relations constrain only the products of the left- and right-handed mixing angles, while the individual mixings are further restricted by electroweak precision observables.

The leading changes to the light-muon $W$ and $Z$ couplings are of order $\theta_{\mu L}^2$. Assuming the electron charged current is unmodified, the modified muon charged current gives
\begin{equation}
\frac{\Gamma(\mu^-\to e^-\bar\nu_e\nu_\mu)}
{\Gamma(\mu^-\to e^-\bar\nu_e\nu_\mu)_{\theta_{\mu L}=0}}
=\cos^2\theta_{\mu L}
\simeq1-\theta_{\mu L}^2,
\qquad G_\mu=G_F\cos\theta_{\mu L}.
\label{eq:muon-lifetime-mixing}
\end{equation}
Since the measured lifetime defines the electroweak input $G_\mu$, it does not by itself give an independent limit on $\theta_{\mu L}$. The constraint must be obtained together with lepton-universality, CKM, $W$-coupling, and $Z$-pole observables~\cite{ParticleDataGroup:2024cfk,ALEPH:2005ab,Crivellin:2020ebi,Poh:2017tfo}. 

The $Z$-pole observables provide a more direct illustration. In the singlet-vector-like-lepton completion, the right-handed $Z\mu\mu$ coupling remains unchanged, while
\begin{equation}
\frac{g_L^{Z\mu}}{g_Z}
=s_W^2-\frac{1}{2}\cos^2\theta_{\mu L}.
\end{equation}
Neglecting lepton masses and flavor-universal radiative corrections, this gives
\begin{equation}
\frac{\Gamma(Z\to\mu^+\mu^-)}
{\Gamma(Z\to e^+e^-)}
=
\frac{\left(s_W^2-\frac12\cos^2\theta_{\mu L}\right)^2+s_W^4}
{\left(s_W^2-\frac12\right)^2+s_W^4}
\simeq1-2.14\theta_{\mu L}^2,
\end{equation}
where $s_W^2=0.231$. The LEP leptonic-width measurements therefore constrain $|\theta_{\mu L}|$ at the few-percent level, approximately
$|\theta_{\mu L}|\lesssim4\times10^{-2}$ at $95\%$ confidence level. Global vector-like-lepton fits can further constrain the relevant charged-current shifts to the per-mille level, depending on the representation and observable set. We therefore adopt the conservative benchmark
\begin{equation}
|\theta_{\mu L,R}|,|\theta_{\tau L,R}|\leq10^{-2},
\label{eq:mixing-benchmark}
\end{equation}
Corrections to the observables considered below are then suppressed by $\mathcal{O}(\theta^2)$ or inverse powers of $M_E$ and can be neglected at the accuracy of our phenomenological recasts.

We next consider the parameters associated with the scalar $S$. For
\begin{equation}
V(S)=-\mu_S^2|S|^2+\lambda_S|S|^4,
\end{equation}
its mass and diagonal lepton couplings in the seesaw-like limit are
\begin{equation}
m_s=\sqrt{2\lambda_S}\,v_s,
\qquad
y_{s\ell\ell}\simeq\frac{m_\ell}{v_s}
=\frac{2g_Xm_\ell}{m_{A'}},
\qquad \ell=\mu,\tau.
\label{y-and-gX}
\end{equation}
Thus, at fixed $(m_{A'},g_X)$, the scalar--lepton couplings are fixed, while $m_s$ is controlled by $\lambda_S$. Equivalently,
\begin{equation}
\lambda_S=2g_X^2\frac{m_s^2}{m_{A'}^2},
\qquad
g_X<\sqrt{\frac{\lambda_S^{\rm max}}{2}}\frac{m_{A'}}{m_s},
\label{eq:scalar-perturbativity-bound}
\end{equation}
so increasing $m_s$ suppresses scalar exchange but eventually becomes limited by perturbativity.

The phenomenological constraints on $s$ depend strongly on its mass. For light scalars, macroscopic-force tests, compact-star dynamics, and stellar cooling can impose stringent limits on $y_{s\mu\mu}$ once loop-induced couplings to photons and ordinary matter are included. Table~\ref{tab:scalar_constraints} lists representative bounds on a light scalar $s$.

\begin{table}[htbp]
\centering
\begin{tabular}{ccc}
\hline
$m_s$ range & representative constraint & approximate $y_{s\mu\mu}^{\rm lim}$ bounds \\
\hline
$m_s<10^{-10}~\mathrm{eV}$ & neutron-star binaries~\cite{Dror:2019uea,Liu:2025zuz} & $\lesssim10^{-20}$ \\
$m_s\lesssim10^{-2}~\mathrm{eV}$ & fifth-force tests~\cite{Hoskins:1985tn,Kapner:2006si,Blakemore:2021zna} & $\lesssim10^{-16}$ \\
$m_s\lesssim100~\mathrm{keV}$ & neutron-star cooling~\cite{Fiorillo:2026wso} & $\lesssim1.2\times10^{-12}$ \\
$m_s\lesssim\mathcal O(10)~\mathrm{MeV}$ & SN~1987A cooling~\cite{Caputo:2021rux,Balazs:2022tjl} & $\lesssim4.2\times10^{-9}$ \\
$100~\mathrm{keV}\lesssim m_s\lesssim\mathcal O(10)~\mathrm{MeV}$ & SN~1987A and diffuse $\gamma$ rays~\cite{Caputo:2021rux,Balazs:2022tjl} & $\lesssim5\times10^{-11}$ \\
\hline
\end{tabular}
\caption{Representative upper bounds $y_{s\mu\mu}\lesssim y_{s\mu\mu}^{\rm lim}(m_s)$ in the low-mass region. The quoted values are order-of-magnitude limits and inherit the production, trapping, and loop-matching assumptions of the cited analyses.}
\label{tab:scalar_constraints}
\end{table}

For any fixed $m_s$, such a coupling bound translates into
\begin{equation}
g_X\lesssim y_{s\mu\mu}^{\rm lim}(m_s)\frac{m_{A'}}{2m_\mu}
\simeq4.7\times10^{-3}\,y_{s\mu\mu}^{\rm lim}(m_s)
\frac{m_{A'}}{\mathrm{MeV}}.
\label{eq:scalar-limit-to-gx}
\end{equation}
The light-scalar searches can therefore provide a powerful complementary probe of the UV completion and, in parts of parameter space, constrain $g_X$ more strongly than direct axial-vector searches.

At intermediate masses, direct searches provide complementary constraints. BaBar and Belle searched for $e^+e^-\to\tau^+\tau^-s$ followed by $s\to e^+e^-,\mu^+\mu^-$ for $0.04~\mathrm{GeV}\lesssim m_s\lesssim6.5$--$7~\mathrm{GeV}$~\cite{BaBar:2020jma,Belle:2022gbl}. Belle~II four-lepton resonance searches reach approximately $9$--$10~\mathrm{GeV}$, but their published interpretations assume a spin-one mediator~\cite{Belle-II:2024wtd,Belle-II:2023ydz}; applying them to $s$ requires a scalar production and acceptance recast. In the mass-proportional benchmark used by BaBar and Belle,
\begin{equation}
\mathcal L\supset-\xi\sum_\ell\frac{m_\ell}{v_h}s\bar\ell\ell,
\qquad
\xi=\frac{v_h}{v_s},
\qquad
g_X<\xi^{\rm lim}(m_s)\frac{m_{A'}}{2v_h}.
\label{eq:scalar-collider-map}
\end{equation}
For example, the Belle limit $\xi^{\rm lim}=0.483$ at $m_s=1~\mathrm{GeV}$ would give $v_s>509~\mathrm{GeV}$, $y_{s\mu\mu}<2.1\times10^{-4}$, and $g_X<9.8\times10^{-4}(m_{A'}/\mathrm{GeV})$ at 90\% confidence level.

These collider limits depend, however, on the visible branching fractions of $s$. In particular, for $m_s>2m_{A'}$ the gauge completion predicts
\begin{equation}
\begin{aligned}
\Gamma(s\to A'A')&=\frac{m_s^3}{32\pi v_s^2}
\left(1-4x+12x^2\right)\sqrt{1-4x},
&x&=\frac{m_{A'}^2}{m_s^2},\\
\Gamma(s\to\ell^+\ell^-)&=\frac{m_\ell^2m_s}{8\pi v_s^2}
\left(1-\frac{4m_\ell^2}{m_s^2}\right)^{3/2}.
\end{aligned}
\label{eq:scalar-partial-widths}
\end{equation}
The $A'A'$ mode can dominate, in which case the visible leptophilic-scalar limits require a branching-ratio recast. Thus the B-factory searches do not provide a universal $g_X$ bound, but they show that the intermediate scalar-mass region cannot simply be assumed unconstrained.

To isolate the axial-vector phenomenology, we therefore choose the scalar-decoupling benchmark $m_s\gtrsim10~\mathrm{GeV}$, 
and assume negligible Higgs-portal mixing. This choice avoids the strongest light-scalar constraints and suppresses scalar exchange, allowing the axial-vector phenomenology to be studied independently to a good approximation. It does not, however, remove the intrinsic correlation between $m_s$ and the axial sector: $y_{s\mu\mu}=2g_Xm_\mu/m_{A'}$ remains independent of $m_s$, while perturbativity requires
\begin{equation}
m_s<\sqrt{2\lambda_S^{\rm max}}\,v_s,
\qquad v_s=\frac{m_{A'}}{2g_X}.
\label{eq:heavy-scalar-existence}
\end{equation}
For example, $m_s=10~\mathrm{GeV}$ and $\lambda_S^{\rm max}=4\pi$ require $v_s\gtrsim2.0~\mathrm{GeV}$. Above the direct B-factory range, there is no comparably broad published scalar exclusion that applies without further assumptions; existing studies mainly provide prospective high-energy sensitivity for larger Yukawa couplings~\cite{Chen:2021leptophilic}. We therefore impose Eq.~\eqref{eq:heavy-scalar-existence} point by point and otherwise treat $s$ as decoupled from the observables below.

In summary, we define the benchmark parameter space by fixing $Q_\ell=1$ and $y_{\mu\tau}=0$, setting $M_{E\mu}=M_{E\tau}\equiv M_E>10~\mathrm{TeV}$ with vanishing kinetic mixing at the matching scale, requiring $|\theta_{\mu L,R}|,|\theta_{\tau L,R}|\leq10^{-2}$, and working in the scalar-decoupling regime $m_s\gtrsim10~\mathrm{GeV}$ with negligible Higgs--singlet mixing. The vector-like-lepton effects are then suppressed by $\mathcal{O}(\theta^2)$ or inverse powers of $M_E$, while scalar-mediated effects and the strong light-scalar constraints are avoided. We additionally impose Eq.~\eqref{eq:heavy-scalar-existence} to ensure perturbative consistency of the scalar sector.

Under these assumptions, the phenomenology considered below is described, to the stated accuracy, by two independent parameters,
\begin{equation}
\{m_{A'},g_X\},
\qquad
v_s=\frac{m_{A'}}{2g_X}.
\label{eq}
\end{equation}
Thus, $v_s$ is fixed by $(m_{A'},g_X)$ rather than varied independently. The remaining UV parameters are not eliminated, but fixed or decoupled by the benchmark assumptions above. Figure~\ref{fig:allconstraints} therefore represents constraints on this conditional two-dimensional slice of the full parameter space; relaxing these assumptions, in particular toward a light scalar, would reintroduce additional mass-dependent constraints such as those summarized in Table~\ref{tab:scalar_constraints}.

\section{Existing Constraints on the Axial-Vector Parameter Space}
\label{sec:phenomenology}
We now turn to the phenomenology of the axial-vector boson. For small light--heavy mixing angles and decoupled vector-like leptons and scalar, the relevant Lagrangian reduces to
\begin{equation}
\begin{aligned}
\mathcal{L}_{A'} \supset & -g_XA'_\alpha\bar\mu\gamma^\alpha\gamma^5\mu
+g_XA'_\alpha\bar\nu_\mu\gamma^\alpha P_L\nu_\mu \\
&+g_X A'_\alpha\bar\tau\gamma^\alpha\gamma^5\tau - g_X A'_\alpha\bar\nu_\tau\gamma^\alpha P_L\nu_\tau,
\end{aligned}
\label{eq:phenomenology-coupling-convention}
\end{equation}
The small vector admixture induced by vector-like lepton mixing is given in Eq.~\eqref{eq:light-muon-Apr}; unless stated otherwise, the phenomenological limits below are quoted in the pure-axial limit. In this limit, the most relevant existing constraints come from neutrino trident production, muon beam-dump searches, invisible charged-meson decays, the muon anomalous magnetic moment, and four-muon resonance searches.

\begin{figure}[t]
\centering
\includegraphics[width=0.8\textwidth]{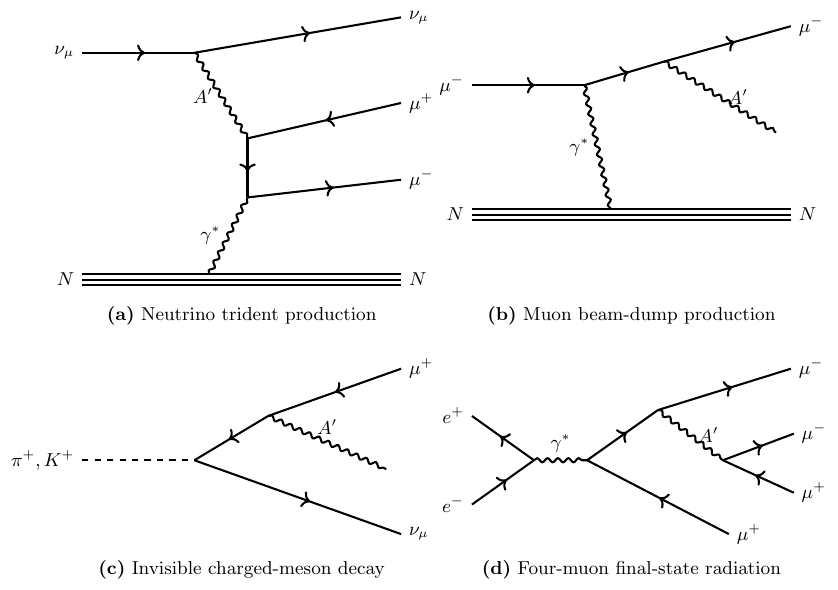}
\caption{Representative $A'$ topologies entering the constraints considered in this work. In panel (a), the $A'$ exchanged from the neutrino line attaches to one vertex on the muon line, while the nuclear photon attaches to the second vertex. In panel (b), the virtual photon transfers momentum to the nucleus and the $A'$ is radiated at a separate vertex on the incoming/outgoing muon line. Panels (c) and (d) show radiation from one charged-muon leg. In panel (c), the diagram with $A'$ emission from the neutrino leg is not displayed but is included in the decay-rate calculation. Crossed diagrams and radiation from the companion charged leg are included where applicable.}
\label{fig:constraint-topologies}
\end{figure}

\subsection{Neutrino Trident Production}
\label{subsec:trident}
Neutrino trident production, $\nu_\mu N\to \nu_\mu N\mu^+\mu^-$, is a direct low-energy probe of the interaction pattern in our model. The CHARM~II and CCFR measurements give
\begin{equation}
\begin{aligned}
\sigma_{\mathrm{CHARM\,II}}/\sigma_{\mathrm{SM}}&=1.58\pm0.57,\\
\sigma_{\mathrm{CCFR}}/\sigma_{\mathrm{SM}}&=0.82\pm0.28,
\end{aligned}
\end{equation}
respectively~\cite{geiregat1990first,CCFR:1991lpl}. Following the equivalent-photon treatment of Refs.~\cite{Williams:1934ad,Belusevic:1987cw,Altmannshofer:2014pba}, the nuclear cross section is obtained by folding the real-photon subprocess with the nuclear photon flux,
\begin{equation}
\sigma(\nu_\mu N\to\nu_\mu N\mu^+\mu^-)
=\int ds\,dq^2\,
\frac{Z^2\alpha}{\pi}\frac{F^2(q^2)}{s\,q^2}\,
\sigma(\nu_\mu\gamma\to\nu_\mu\mu^+\mu^-),
\label{eq:trident-epa}
\end{equation}
where $Ze$ and $F(q^2)$ are the nuclear charge and electromagnetic form factor. The SM coefficients are $C_V=1/2+2s_W^2$ and $C_A=1/2$. In the massless-muon limit, the axial mediator contributes to the axial coefficient through the diagram shown in panel (a) of Fig.~\ref{fig:constraint-topologies}, giving
\begin{equation}
C_A\longrightarrow C_A-\frac{g_X^2}{\sqrt2G_F}\frac{1}{k^2-m_{A'}^2},
\end{equation}
where $k$ is the momentum carried by the exchanged $A'$. This differs from the conventional vector $L_\mu-L_\tau$ model, which shifts $C_V$. In the heavy-mediator limit the ratio reduces to
\begin{equation}
\frac{\sigma^{\rm SM+A'}}{\sigma^{\rm SM}}
\simeq
\frac{C_V^2+\left[C_A+g_X^2/(\sqrt2G_Fm_{A'}^2)\right]^2}
{C_V^2+C_A^2}.
\end{equation}
At leading order in the new interaction, the method of Ref.~\cite{Altmannshofer:2014pba} can be reused directly at the level of the SM--new-physics interference. For a vector $L_\mu-L_\tau$ mediator the interference is proportional to $C_V\delta C_V$, whereas in the present model one makes the replacement
\begin{equation}
C_V\,\delta C_V^{(Z')}(k^2)
\longrightarrow
C_A\,\delta C_A^{(A')}(k^2),
\qquad
\delta C_A^{(A')}(k^2)=
\frac{g_X^2}{\sqrt2G_F\left(m_{A'}^2-k^2\right)}.
\label{eq:trident-leading-substitution}
\end{equation}
This simple substitution is useful for checking the sign, normalization, and heavy-mediator behavior of the axial result.

We do not obtain the axial contours by digitizing the vector bound in Fig.~2 of Ref.~\cite{Altmannshofer:2014pba}. For the final limits, we evaluate the full spin- and polarization-summed Bethe--Heitler amplitude for $\nu_\mu\gamma\to\nu_\mu\mu^+\mu^-$, retain finite $m_\mu$, integrate the three-body phase space numerically, and perform the EPA convolution with the corresponding experimental fluxes. The olive and yellow curves in Fig.~\ref{fig:allconstraints} are the resulting CCFR and CHARM~II bounds, respectively; CCFR gives the stronger limit over most of their common mass range.

\subsection{Muon Beam-Dump Searches}
\label{subsec:muon-beam-dump}

Muon beam-dump experiments provide a complementary direct probe of the charged-muon interaction. Ref.~\cite{Chakraborty:2025jbd} studied the present and projected sensitivities of NA64$_\mu$~\cite{NA64:2024nwj}, M$^3$~\cite{Kahn:2018cqs}, MuSIC~\cite{Acosta:2021qpx,Acosta:2022ejc}, and a future muon beam dump~\cite{Cesarotti:2022ttv} to a vector $U(1)_{L_\mu-L_\tau}$ gauge boson. To obtain an approximate NA64$_\mu$ constraint for the axial interaction, we follow the rate-matching procedure of Ref.~\cite{Chakraborty:2025jbd}.
For kinematically allowed $A'$, the most relevant production channel is
\begin{equation}
\mu^- A(\text{Pb}) \to \mu^- A(\text{Pb}) A', \quad A' \to \text{invisible}
\label{eq:NA64production}
\end{equation}
where $A(\mathrm{Pb})$ denotes a lead target nucleus~\cite{andreev2024exploration}, and the axial vector is emitted from the incoming or outgoing muon line. Retaining the finite-muon-mass terms, the bremsstrahlung spectrum is~\cite{Chakraborty:2025jbd}
\begin{equation}
\begin{aligned}
\small
\frac{d\sigma^{A'}_{2\to 3}}{dx}
&=\frac{\alpha^2\chi\,\beta\,g_X^2 x}{3\pi m_{A'}^2\left[m_{A'}^2(1-x)+m_\mu^2x^2\right]^2} \times\left[ 6m_\mu^4x^4 -2m_{A'}^4(x-1)\{x(x-3)+3\} +m_\mu^2m_{A'}^2x^2(x-4)(3x-4) \right],
\label{eq:beamdump-axial-spectrum}
\end{aligned}
\end{equation}
where $x=E_{A'}/E_\mu$, $\chi$ is the effective photon flux of the target, and $\beta$ is the final-state velocity factor; further details are given in Ref.~\cite{Chakraborty:2025jbd}. The axial production rate coincides with the vector result in the formal $m_\mu\to0$ limit. At low $m_{A'}$, axial-current nonconservation enhances longitudinal $A'$ emission, and the production rate scales approximately as $m_\mu^2/m_{A'}^2$ relative to the vector benchmark.

The expected number of NA64$_\mu$ signal events for invisible decays $A'\to\nu\bar\nu$ can be written as
\begin{equation}
N_{\text{sig}} = \frac{N_{\text{MOT}} n_{\text{Pb}}}{\langle dE_\mu/dy \rangle} \int_{E_{\mu,\min}}^{E_{\mu,\max}} dE_0 \int_{x_{\min}}^{x_{\max}} dx \frac{d\sigma_{2\to3}^{A'}}{dx} Br(A' \to \bar{\nu}\nu) N(m_{A'})
\label{eq:NA64Signal}
\end{equation}
where $N_{\text{MOT}}$ is the number of muons on target, $n_{\text{Pb}}$ is the number density of lead nuclei, $\langle dE_\mu/dy\rangle$ is the average muon energy loss per unit length~\cite{navas2024review}, and $\mathrm{Br}(A'\to\nu\bar\nu)$ is the invisible branching fraction. The effective mass-dependent factor $N(m_{A'})$ is calibrated by reproducing the current four-dimensional NA64$_\mu$ result in Fig.~3 of Ref.~\cite{andreev2024exploration}, obtained with $N_{\text{MOT}}\simeq2\times10^{10}$. We adopt the integration ranges specified in Secs.~III and IV of Ref.~\cite{Chakraborty:2025jbd}. In the zero-background Poisson approximation, a 90\% confidence-level exclusion corresponds to $N_{\text{sig}}=2.3$. Rather than performing an independent detector simulation, we obtain the axial contour by matching its event yield to the published vector exclusion of Ref.~\cite{Chakraborty:2025jbd}:
\begin{equation}
g_X^{\rm lim}(m_{A'})
=g_{\rm vec}^{\rm lim}(m_{A'})
\left[
\frac{N_{\text{sig}}^V(g'=1)}{N_{\text{sig}}^A(g'=1)}
\right]^{1/2},
\label{eq:beamdump-rescale}
\end{equation}
The result is shown in Fig. \ref{fig:allconstraints}. Since the curves from Ref.~\cite{Chakraborty:2025jbd} are digitized from their $n_{\rm KKmax}=1$, $\epsilon_4=\widetilde y_{\rm SM}=0$ figure and then rescaled only at the production-rate level, the plot should be read as an order-of-magnitude phenomenological projection rather than as an official experimental limit. A robust contour would require the experiment-specific target geometry, missing-energy acceptance, finite-lifetime effects, detector response, digitization uncertainty, and the range of validity of the improved Weizsacker-Williams approximation to be propagated together.

\subsection{Invisible Meson Decays}
\label{subsec:pion-invisible}

Stopped-pion experiments provide an additional low-mass handle on the same axial interaction. Ref.~\cite{Altmannshofer:2026opc} studied the invisible three-body decays $\pi^+\to \ell^+\nu_\ell X$ for a scalar or vector $X$ and recast the PIENU search~\cite{PIENU:2021clt}. We show the resulting PIENU bound in Fig.~\ref{fig:allconstraints}. For this $U_A(1)_{L_\mu-L_\tau}$ model, the most relevant channel is
\begin{equation}
    \pi^+\to \mu^+ \nu_\mu +A'
\end{equation}
with $A'$ radiated from the external muon and muon-neutrino legs. The decay is open only for $m_{A'}<m_{\pi^+}-m_\mu\simeq33.9~{\rm MeV}$. The electron channel considered in Ref.~\cite{Altmannshofer:2026opc} is absent at tree level in our model. Moreover, the parity-even photon--$A'$ kinetic mixing vanishes at one loop in the pure-axial limit; any electronic coupling induced by the small vector admixture or by the heavy sector is benchmark dependent and suppressed. Throughout the pion-decay window, $m_{A'}<33.9~\mathrm{MeV}<2m_\mu$, so the charged-muon decay mode is closed while the neutrino modes remain open. We therefore treat $A'$ as invisible in the PIENU recast.

For our pure-axial benchmark, the normalized differential rate can be written as \cite{Altmannshofer:2026opc}
\begin{equation}\label{eq:pion-axial-spectrum}
\small
\begin{aligned}
\frac{d{\rm BR}(\pi^+\to\mu^+\nu_\mu A')}
{{\rm BR}(\pi^+\to\mu^+\nu_\mu)}
&=\frac{g_X^2}{2\pi^2}\,
\frac{dE_\mu dE_{A'}}{m_\pi^2}\,
\frac{1}{(1-x_\mu)^2}\cdot \left[\frac{x_{\mu A'}+2-3 x_{\mu}}{x_{\nu A'}}-\frac{2\left(x_{\mu A'}-1+x_{\nu A'}\right)}{x_{A'}}-\frac{\left(1-x_{\mu}\right) x_{A'}}{x_{\nu A'}^2}\right.\\
& \left. +\frac{2 x_{\mu}}{x_{\mu A'}-x_{\mu}}\left(\frac{x_{A'}}{x_{\nu A'}}-\frac{x_{\nu A'}}{x_{A'}}\right) -\frac{2\left(1-x_{\mu}\right)^2}{\left(x_{\mu A'}-x_{\mu}\right) x_{\nu A'}}+\frac{2+x_{\nu A'}-6 x_{\mu}}{x_{\mu A'}-x_{\mu}}-\frac{\left(1-x_{\mu}\right)\left(x_{A'}-4 x_{\mu}\right)}{\left(x_{\mu A'}-x_{\mu}\right)^2}\right],
\end{aligned}
\end{equation}
where $x_{\mu/A'}=m_{\mu/A'}^2/m_\pi^2$ and $x_{ab}=(p_a+p_b)^2/m_\pi^2$, as defined in Ref.~\cite{Altmannshofer:2026opc}. This spectrum distorts the low-energy tail of the muon kinetic-energy distribution rather than producing a narrow line. Applying the fitting procedure of Ref.~\cite{Altmannshofer:2026opc} to the PIENU muon-energy spectrum gives the limit shown in Fig.~\ref{fig:allconstraints}. Present PIENU data exclude approximately $g_X\gtrsim{\rm few}\times10^{-3}$ at MeV masses, with the sensitivity weakening to $\mathcal O(10^{-1})$ as $m_{A'}$ approaches the pion-decay endpoint.

The analogous kaon decay has the larger kinematic range
\begin{equation}
K^+\to\mu^+\nu_\mu A',
\qquad
m_{A'}<m_{K^+}-m_\mu\simeq388.0~\mathrm{MeV}.
\end{equation}
This endpoint is well above the dimuon threshold, $2m_\mu\simeq211.3~\mathrm{MeV}$, so the interval in which $A'\to\mu^+\mu^-$ is open is substantial. Below the tau threshold and neglecting loop-induced electronic decays,
\begin{equation}
\Gamma(A'\to\nu\bar\nu)=\frac{g_X^2m_{A'}}{12\pi},
\qquad
\Gamma(A'\to\mu^+\mu^-)=\frac{g_X^2m_{A'}}{12\pi}
\left(1-\frac{4m_\mu^2}{m_{A'}^2}\right)^{3/2},
\end{equation}
where the neutrino width is summed over $\nu_\mu$ and $\nu_\tau$. Hence
\begin{equation}
{\rm Br}_{\rm inv}(A')=
\left[1+\Theta(m_{A'}-2m_\mu)
\left(1-\frac{4m_\mu^2}{m_{A'}^2}\right)^{3/2}\right]^{-1}.
\label{eq:kaon-Apr-invisible-branching}
\end{equation}
NA62 reports mass-dependent limits on three-body $K^+\to\mu^+\nu X$ decays for specific scalar and vector templates, as well as ${\rm Br}(K^+\to\mu^+\nu\bar\nu\nu)<1.0\times10^{-6}$ for the SM four-body spectrum~\cite{NA62:2021bji}. Neither result is a mass-independent limit on the axial signal considered here. The teal curve in Fig.~\ref{fig:allconstraints} shows the adopted axial reinterpretation, including the invisible branching fraction in Eq.~\eqref{eq:kaon-Apr-invisible-branching}. Because a complete detector-level acceptance map for the axial spectrum is unavailable, this curve should be regarded as an indicative recast rather than an official NA62 limit.

\subsection{Muon $g-2$}
\label{sec:muon_g-2}

The muon anomalous magnetic moment is a particularly sharp diagnostic of the Lorentz structure of the new force. The current experimental world average, dominated by the final Fermilab E989 result, is $a_\mu^{\rm exp}=116592071.5(14.5)\times10^{-11}$ \cite{Muong-2:2025xyk}. Using the 2025 Muon $g-2$ Theory Initiative Standard-Model update based on lattice-QCD calculations gives $\Delta a_\mu\equiv a_\mu^{\rm exp}-a_\mu^{\rm SM} =38(63)\times10^{-11}$ \cite{Muong-2:2025xyk,Aliberti:2025beg}. This is no longer a robust positive anomaly in the same sense as the older data-driven WP20 comparison \cite{Aoyama:2020ynm}; therefore, we use $(g-2)_\mu$ mainly as a consistency constraint on the axial model.

Generally, for any new neutral spin-one boson with
\begin{equation}
\mathcal L\supset
A'_\alpha\bar\mu\gamma^\alpha
\left(g_\mu^V-g_\mu^A\gamma_5\right)\mu ,
\end{equation}
the standard one-loop result can be written as \cite{Leveille:1977rc,Davoudiasl:2014kua,Anastasopoulos:2022ywj}
\begin{equation}
\Delta a_\mu^{A'}=
\frac{m_\mu^2}{4\pi^2m_{A'}^2}
\left[
(g_\mu^V)^2F_V(r)
-(g_\mu^A)^2F_A(r)
\right],
\qquad
r\equiv \frac{m_\mu^2}{m_{A'}^2},
\label{eq:gminus2-vector-axial}
\end{equation}
where
\begin{equation}
F_V(r)=\int_0^1 dx\,
\frac{x^2(1-x)}{1-x+r x^2},
\qquad
F_A(r)=\int_0^1 dx\,
\frac{x(1-x)(4-x)+2r x^3}{1-x+r x^2}.
\label{eq:gminus2-FV-FA}
\end{equation}
The relative minus sign is the main difference from a vector $L_\mu-L_\tau$ gauge boson. In the heavy-mediator limit,
\begin{equation}
\Delta a_\mu^{A'}\simeq
\frac{m_\mu^2}{12\pi^2m_{A'}^2}
\left[(g_\mu^V)^2-5(g_\mu^A)^2\right],
\qquad m_{A'}\gg m_\mu ,
\label{eq:gminus2-heavy-limit}
\end{equation}
whereas for a light mediator the vector contribution approaches the Schwinger form $+(g_\mu^V)^2/(8\pi^2)$, while the axial contribution is enhanced by the longitudinal polarization,
\begin{equation}
\Delta a_\mu^{A'}\simeq
\frac{(g_\mu^V)^2}{8\pi^2}
-\frac{(g_\mu^A)^2}{4\pi^2}\frac{m_\mu^2}{m_{A'}^2},
\qquad m_{A'}\ll m_\mu .
\label{eq:gminus2-light-limit}
\end{equation}
For the reduced effective Lagrangian defined in Eq.~\eqref{eq:phenomenology-coupling-convention}, $g_\mu^V=0$ and $g_\mu^A=g_X$, so the axial-vector correction is negative. The scalar in the UV completion gives the positive contribution~\cite{Leveille:1977rc,Tucker-Smith:2010wdq}
\begin{equation}
\Delta a_\mu^s=
\frac{y_{s\mu\mu}^2}{8\pi^2}
\int_0^1dx\,
\frac{x^2(2-x)m_\mu^2}
{m_s^2(1-x)+x^2m_\mu^2}
=\frac{g_X^2}{2\pi^2}\frac{m_\mu^2}{m_{A'}^2}
\int_0^1dx\,
\frac{x^2(2-x)m_\mu^2}
{m_s^2(1-x)+x^2m_\mu^2},
\label{eq:gminus2-scalar}
\end{equation}
where Eq.~\eqref{y-and-gX} was used in the second equality. The complete result is therefore
\begin{equation}
\Delta a_\mu=\Delta a_\mu^{A'}+\Delta a_\mu^s.
\label{eq:gminus2-total}
\end{equation}
For $m_s\ll m_\mu$,
\begin{equation}
\Delta a_\mu^s\simeq
\frac{3g_X^2}{4\pi^2}\frac{m_\mu^2}{m_{A'}^2},
\end{equation}
which is three times the magnitude of the light-$A'$ axial limit in Eq.~\eqref{eq:gminus2-light-limit}. The scalar then dominates and the total correction is positive. As $m_s$ increases, the positive term decreases and can cancel the axial contribution for $m_s=\mathcal O(m_\mu)$; the precise cancellation mass depends on $m_{A'}$. For still larger $m_s$, the scalar loop decouples as $m_\mu^2/m_s^2$ up to a logarithm, leaving the negative axial-vector contribution.

\begin{figure}[htbp]
    \centering
    \includegraphics[width=0.48\textwidth]{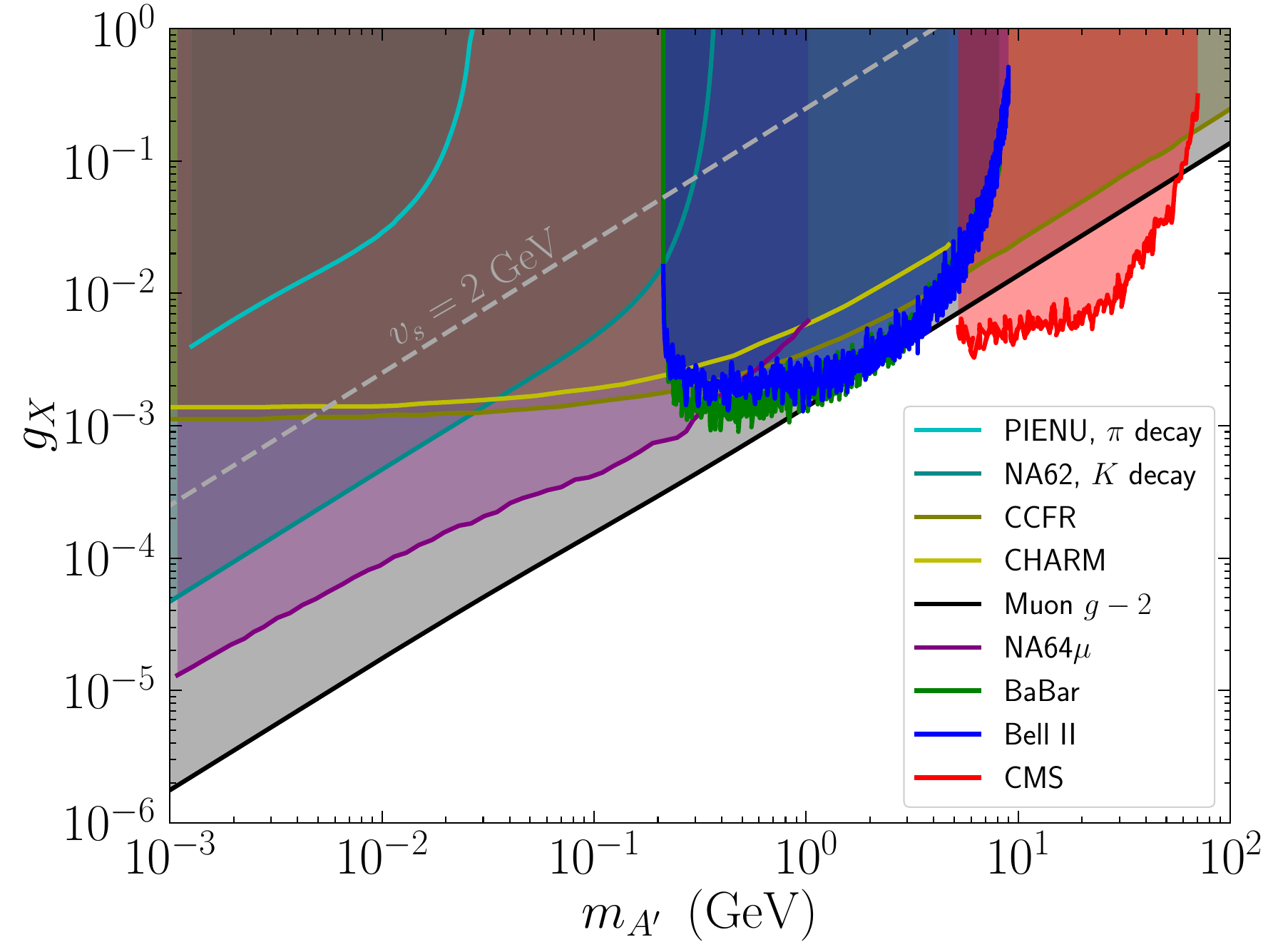}\hfill
    \includegraphics[width=0.48\textwidth]{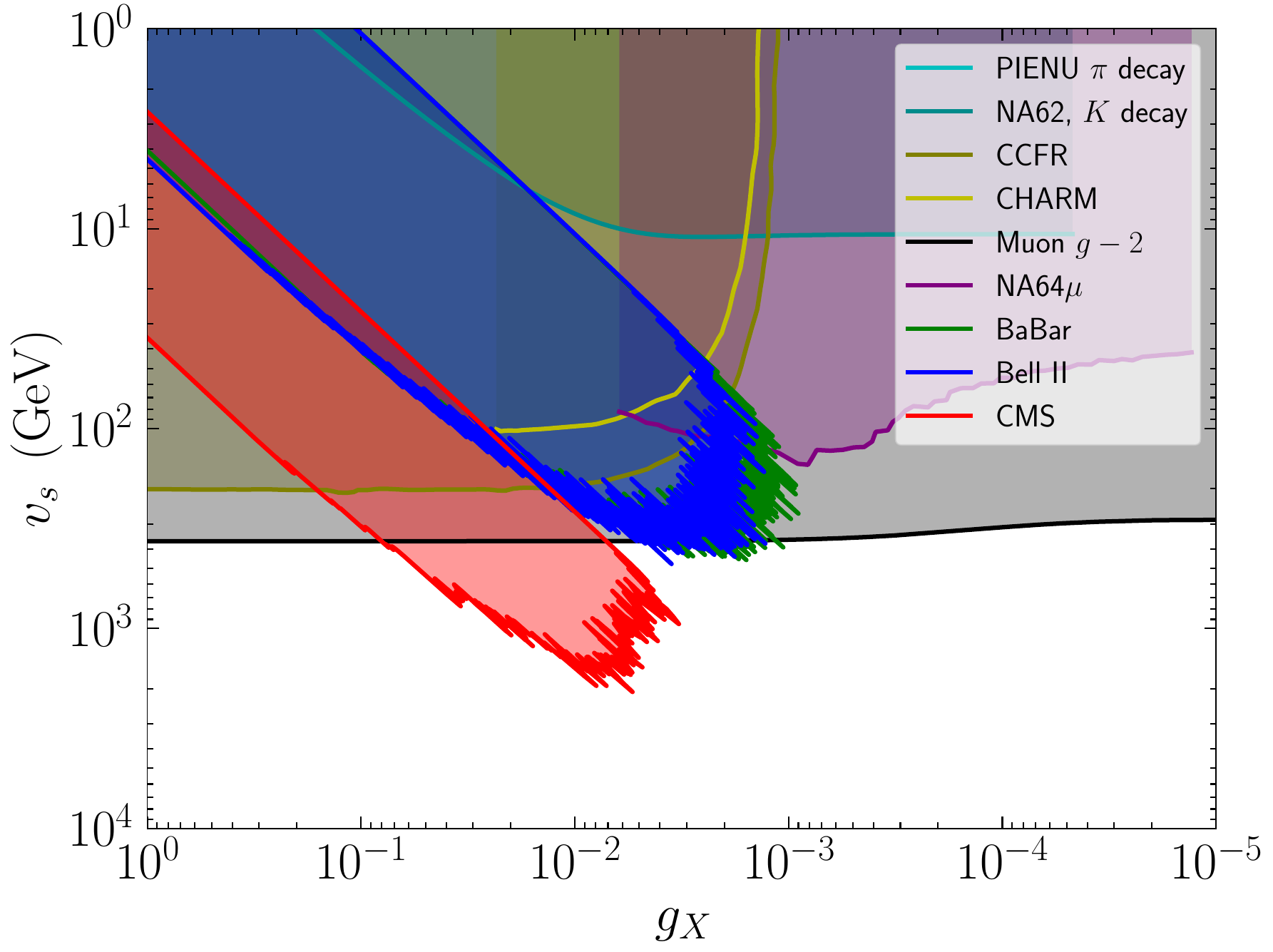}
    \caption{Existing constraints on the pure-axial effective-current benchmark. Left: excluded regions in the $(m_{A'},g_X)$ plane from PIENU pion decay (cyan), the indicative NA62 kaon-decay recast (teal), CCFR and CHARM~II neutrino trident production (olive and yellow), the scalar-decoupled $(g-2)_\mu$ constraint (black), the NA64$_\mu$ recast (purple), and the BaBar, Belle~II, and CMS four-muon searches (green, blue, and red). The gray dashed line marks $v_s=m_{A'}/(2g_X)=2~\mathrm{GeV}$; for the benchmark $m_s\geq10~\mathrm{GeV}$ with $\lambda_S\leq4\pi$, the region above this line does not satisfy perturbativity. Right: the same constraints mapped to the $(g_X,v_s)$ plane using $v_s=m_{A'}/(2g_X)$; note that $g_X$ decreases from left to right and $v_s$ increases from top to bottom. The right panel contains no additional likelihood information. The mass-dependent scalar limits in Table~\ref{tab:scalar_constraints} are not superimposed.}
    \label{fig:allconstraints}
\end{figure}

The black $(g-2)_\mu$ contour in Fig.~\ref{fig:allconstraints} is therefore not valid for arbitrary $m_s$: it corresponds to the scalar-decoupling benchmark $m_s\gtrsim10~\mathrm{GeV}$ specified above. In this benchmark the scalar term is below $0.2\%$ of the pure-axial contribution, and the approximate $2\sigma$ consistency condition from the 2025 comparison is $\Delta a_\mu^{A'}\gtrsim-8.8\times10^{-10}$. For a light scalar, Eq.~\eqref{eq:gminus2-total} must instead be used and the exclusion can weaken or disappear near the cancellation region.

Notably, in the low-mass regime, the pure-axial $(g-2)_\mu$ exclusion contour is considerably tighter than the conventional vector $L_\mu-L_\tau$ contour because of the longitudinal contribution associated with axial-current nonconservation~\cite{Fayet:2020nyx,Dror:2017nsg,Biggio:2016wyy}. With nonzero mixing, Eq.~\eqref{eq:light-muon-Apr} gives $g_\mu^V=g_X\sin^2\theta_{\mu R}$ and $g_\mu^A=-g_X\cos^2\theta_{\mu R}$. Under Eq.~\eqref{eq:mixing-benchmark}, the induced vector term and vector-like-lepton loop effects are negligible.

The left panel of Fig.~\ref{fig:allconstraints} shows how the dominant probe changes with mediator mass. PIENU is restricted to $m_{A'}<m_\pi-m_\mu$, while the kaon channel extends the meson-decay reach to $m_{A'}<m_K-m_\mu$; the sharp deterioration of both curves near their endpoints is a phase-space effect. CCFR and CHARM~II give broad trident constraints: their limits are nearly mass independent for a light mediator and weaken when the exchange becomes contact-like. NA64$_\mu$ supplies the strongest direct missing-energy reach over much of the sub-GeV region. Once $A'\to\mu^+\mu^-$ opens, BaBar and Belle~II probe resonant final-state radiation up to the several-GeV range, while CMS extends the same visible strategy to masses of order tens of GeV. The black $(g-2)_\mu$ curve is numerically the strongest bound over much of the low-mass plane because the longitudinal mode enhances the negative axial contribution. It is qualitatively different from the production bounds: it assumes the heavy-scalar benchmark, and it can move or disappear if a light scalar cancels the $A'$ loop.

The right panel displays exactly the same exclusions in terms of the symmetry-breaking scale. With the plotted axis convention, moving right means decreasing $g_X$, while moving downward means increasing $v_s$. The nearly horizontal black boundary reflects the approximate low-mass scaling $g_X^{\rm lim}\propto m_{A'}$ of the axial $(g-2)_\mu$ constraint, which translates into an approximately fixed $v_s=m_{A'}/(2g_X)$. The $v_s=2~\mathrm{GeV}$ guide in the left panel highlights the region where the symmetry-breaking scale is small and $y_{s\mu\mu}=m_\mu/v_s$ becomes large. It is not itself an experimental bound. Consequently, Fig.~\ref{fig:allconstraints} summarizes constraints on the chosen effective-current slice; the heavy-scalar consistency condition and the light-scalar bounds must still be applied separately when embedding a plotted point into the UV model.

\subsection{$\mu^+\mu^-$ resonance search in four-muon final state}

The axial-vector boson $A'$ can also be produced through final-state radiation (FSR) from the $\mu^+\mu^-$ pair in electron--positron annihilation~\cite{BaBar:2016sci, Belle-II:2022yaw, Belle-II:2024wtd} or in the decay of an on-shell $Z$ boson~\cite{Harigaya:2013twa, delAguila:2014soa, CMS:2018yxg}. In the parameter region of interest, the produced $A'$ promptly decays into a pair of charged leptons or neutrinos, leading to either a four-charged-lepton final state or a missing-energy signature. In this work, we focus on the four-muon channels,
\begin{equation}
	\begin{aligned}
		e^+e^- &\rightarrow \mu^+\mu^-A', \qquad A'\rightarrow \mu^+\mu^-,\\
		Z &\rightarrow \mu^+\mu^-A', \qquad A'\rightarrow \mu^+\mu^-.
	\end{aligned}
\end{equation}

The searches for these processes require the four-muon invariant mass to be consistent with the center-of-mass energy of the $e^+e^-$ collision or the $Z$ boson mass, respectively. In addition, the resonant decay $A'\to\mu^+\mu^-$ can be used to further reduce the background. Consequently, the constraints obtained from the four-muon final state are generally stronger than those from missing-energy searches based on $A'\rightarrow\nu\bar{\nu}$~\cite{Belle-II:2022yaw}. Therefore, throughout this work we consider only the limits derived from the four-muon searches.

For sufficiently heavy axial-vector bosons, the difference between vector and axial-vector FSR arises only from the longitudinal polarization of the emitted boson. This contribution is proportional to $m_\mu^2/m_{A'}^2$ and rapidly becomes negligible for $m_{A'}\gtrsim m_\mu$. As a result, the differential cross sections of vector and axial-vector radiation are nearly identical, leading to almost identical selection efficiencies after the experimental cuts. The experimental limits on the axial $U_A(1)_{L_\mu-L_\tau}$ model can therefore be obtained by rescaling the published limits on the corresponding vector model according to
\begin{equation}
	\begin{aligned}
		g_X^{\rm lim}(m_{A'})
		&=
		g_{\rm vec}^{\rm lim}(m_{A'})
		\left[
		\frac{\sigma_V^{\rm FSR}(g_V=1)\,
			{\rm Br}_V(\mu^+\mu^-)}
		{\sigma_A^{\rm FSR}(g_X=1)\,
			{\rm Br}_A(\mu^+\mu^-)}
		\right]^{1/2}
		\simeq
		g_{\rm vec}^{\rm lim}(m_{A'})
		\left[
		\frac{{\rm Br}_V(\mu^+\mu^-)}
		{{\rm Br}_A(\mu^+\mu^-)}
		\right]^{1/2},
		\\[2mm]
		g_X^{\rm lim}(m_{A'})
		&=
		g_{\rm vec}^{\rm lim}(m_{A'})
		\left[
		\frac{\Gamma_V^{\rm FSR}(g_V=1)\,
			{\rm Br}_V(\mu^+\mu^-)}
		{\Gamma_A^{\rm FSR}(g_X=1)\,
			{\rm Br}_A(\mu^+\mu^-)}
		\right]^{1/2}
		\simeq
		g_{\rm vec}^{\rm lim}(m_{A'})
		\left[
		\frac{{\rm Br}_V(\mu^+\mu^-)}
		{{\rm Br}_A(\mu^+\mu^-)}
		\right]^{1/2},
	\end{aligned}
	\label{eq:four-muon-rescale}
\end{equation}
where the subscripts $A$ and $V$ denote quantities in the axial and vector $U(1)_{L_\mu-L_\tau}$ models, respectively. Here $\sigma^{\rm FSR}$ denotes the cross section for $e^+e^-\rightarrow\mu^+\mu^-A'$, while $\Gamma^{\rm FSR}$ denotes the decay width for $Z\rightarrow\mu^+\mu^-A'$. The branching fraction for $A'\rightarrow\mu^+\mu^-$ is denoted by ${\rm Br}(\mu^+\mu^-)$, and $g_{\rm vec}^{\rm lim}$ represents the published exclusion limit on the gauge coupling in the vector $U(1)_{L_\mu-L_\tau}$ model.

We reinterpret the BaBar~\cite{BaBar:2016sci}, Belle~II~\cite{Belle-II:2024wtd}, and CMS~\cite{CMS:2018yxg} results using the same rate-matching principle, but not an identical production calculation. For BaBar and Belle~II the matched quantity is $\sigma(e^+e^-\to\mu^+\mu^-A')\,{\rm Br}(A'\to\mu^+\mu^-)$, whereas for CMS it is $\Gamma(Z\to\mu^+\mu^-A')\,{\rm Br}(A'\to\mu^+\mu^-)$. The approximation that vector and axial acceptances are equal is controlled only for $m_{A'}\gg m_\mu$; near threshold these reinterpretations should be regarded as rate-level estimates pending detector-level simulations. We therefore discuss them as complementary direct searches but do not combine them statistically with the low-energy contours in Fig.~\ref{fig:allconstraints}.

\subsection{Solar Neutrino}
Because $A'$ couples to left-handed neutrinos, solar muon- and tau-neutrino currents can source a coherent background field~\cite{Fang:2025fvo}. This background changes the muon spin-precession signal and can therefore be constrained with $(g-2)_\mu$ data.

We consider solar neutrinos as the dominant source, since they provide the largest neutrino flux at the detector location. The muon- and tau-neutrino components arise through neutrino oscillations during propagation from the Sun to the Earth.
In the relativistic limit, the left-handed neutrino current acts as a source for a Yukawa field. Crucially, the source is weighted by the flavor charges. In the convention $Q_e=0$, $Q_\mu=+1$, and $Q_\tau=-1$, the field is
\begin{equation}
A'^{\alpha}(\vec{x}_d)
=\int d^3x\,g_X
\sum_{\ell=e,\mu,\tau}Q_\ell\Phi_{\nu_\ell}^{\alpha}(x)
\frac{e^{-m_{A'}\Delta r}}{4\pi\Delta r}
=\int d^3x\,g_X
\left(\Phi_{\nu_\mu}^{\alpha}-\Phi_{\nu_\tau}^{\alpha}\right)
\frac{e^{-m_{A'}\Delta r}}{4\pi\Delta r},
\label{eq:axial-vector-potential}
\end{equation}
where $x_d$ is the detector position, $\Phi_{\nu_\ell}^{\alpha}$ is the flavor-resolved neutrino current, and $\Delta r=|\vec{x}_d-\vec{x}|$. Equivalently, for flavor fractions $f_\ell$, the effective charge per solar neutrino is
\begin{equation}
q_\nu^{\rm eff}=\sum_\ell Q_\ell f_\ell=f_\mu-f_\tau.
\end{equation}
Equal muon- and tau-neutrino fluxes therefore cancel; the signal is controlled by the flavor asymmetry generated by oscillations. The induced field gives a characteristic day--night sign change on top of a small time-independent offset.

Using the day--night modulation estimate of Ref.~\cite{Fang:2025fvo}, in the regime where the interaction range $m^{-1}_{A'}$ is larger than the distance between the Sun and Earth, the potential approaches that of a massless mediator and gives the constraint
\begin{equation}
g_X < 5.8 \times 10^{-18}\  (\text{massless limit}). 
\end{equation}
For larger mediator masses ($m_{A'} > 10^{-17}$ eV), the potential scales as $m^{-2}_{A'}$. The same day--night modulation analysis gives
\begin{equation}
g_X < 3.1 \times   \frac{ m_{A'}}{{\rm eV}}\ (m_{A'} > 10^{-17}\ \rm{eV}). 
\end{equation}
This long-range-force constraint applies to ultralight mediators, far below the MeV--TeV region displayed in Fig.~\ref{fig:allconstraints}, and is therefore not shown there.

\section{Future Muon Collider Searches}
\label{sec:muon-collider}

A high-energy muon collider would probe the same muon-philic force in a very different mass range. Ref. \cite{Huang:2021nkl} studied the ordinary vector $L_\mu-L_\tau$ gauge boson at a benchmark $\sqrt{s}=3~{\rm TeV}$ muon collider, using both the two-body process $\mu^+\mu^-\to\ell^+\ell^-$ and an initial photon emission process $\mu^+\mu^-\to\gamma A'$. 
The two-body channel is particularly sensitive to off-shell and near-resonance distortions of the angular distribution, whereas radiative return provides the leading on-shell strategy for $m_{A'}<\sqrt{s}$. In the ultrarelativistic limit, the inclusive radiative-return rate is only mildly sensitive to whether the charged-lepton current is vector or axial. We therefore focus on the ability of angular observables in the two-body channel to distinguish the two interaction structures. A comprehensive detector-level study would closely follow the framework of Ref.~\cite{Huang:2021nkl} and is beyond the scope of the present work.
\begin{figure}[htbp]
\centering
\includegraphics[width=0.94\textwidth]{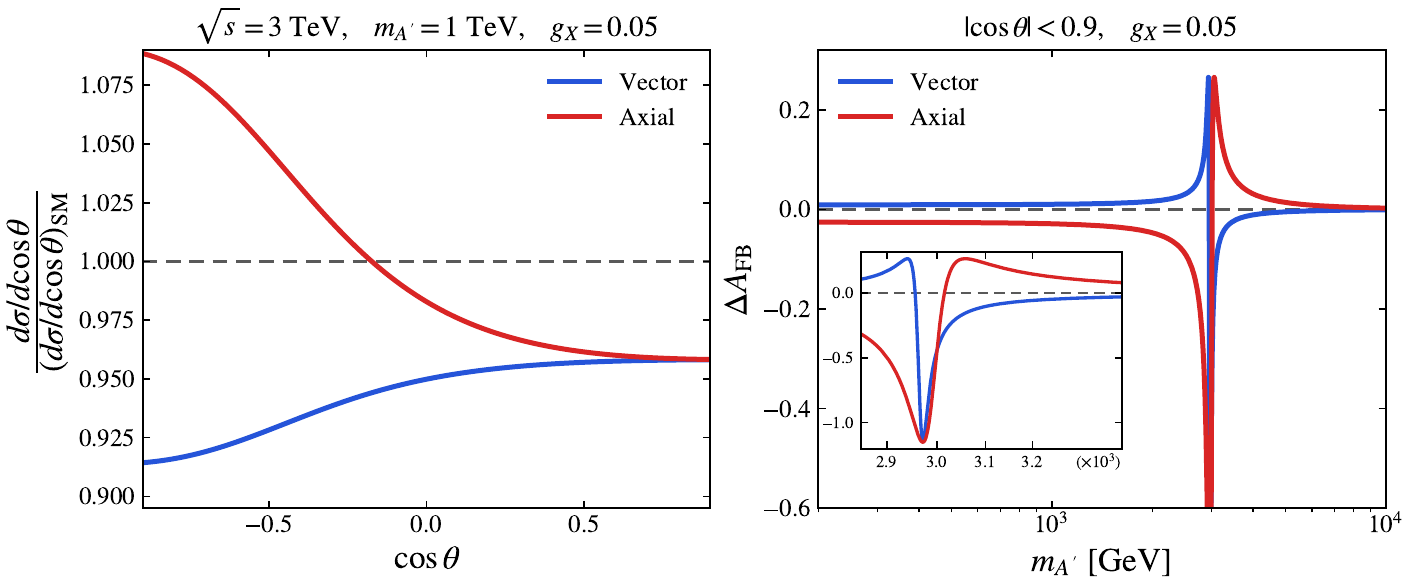}
\caption{Future muon-collider angular comparison for $\mu^+\mu^-\to\tau^+\tau^-$ at $\sqrt{s}=3~{\rm TeV}$ and $g_X=0.05$. Left: ratio of the differential cross section to the SM prediction for $m_{A'}=1~{\rm TeV}$. Right: shift in $A_{\rm FB}$, integrated over $|\cos\theta|<0.9$, as a function of $m_{A'}$. The main panel uses a restricted vertical range to resolve the nonresonant behavior, while the inset displays the full finite-width structure around $m_{A'}\simeq\sqrt{s}$. We use $\Gamma_{A'}=g_X^2m_{A'}/(4\pi)$, which gives $\Gamma_{A'}\simeq0.60~\mathrm{GeV}$ at the resonance. The calculation uses the massless external-lepton chiral-amplitude formula in Eq.~\eqref{eq:mucollider-chiral-xsec}.}
\label{fig:muon-collider-angular}
\end{figure}

For the two-body process $\mu^+\mu^-\to\ell^+\ell^-$, the ditau final state receives only $s$-channel $\gamma,Z,A'$ exchange and therefore gives the cleanest analytic illustration of the axial interference. 
Neglecting external lepton masses, the unpolarized differential cross section can be written in terms of chiral amplitudes as
\begin{align}
\frac{d\sigma_{\ell\ell}}{d\cos\theta}=
\frac{s}{128\pi}
\bigg[ \left(|\mathcal A_{LL}|^2+|\mathcal A_{RR}|^2\right)(1+\cos\theta)^2 + \left(|\mathcal A_{LR}|^2+|\mathcal A_{RL}|^2\right)(1-\cos\theta)^2 \bigg],
\label{eq:mucollider-chiral-xsec}
\end{align}
with
\begin{equation}
\mathcal A_{ij}
=\frac{e^2}{s}+\frac{g_i^Z g_j^Z}{s-m_Z^2+i m_Z\Gamma_Z}+\frac{g_i^\mu g_j^\tau}{s-m_{A'}^2+i m_{A'}\Gamma_{A'}},
\qquad i,j=L,R .
\label{eq:mucollider-chiral-amplitude}
\end{equation}
We can see that, the coupling combinations $g_i^\mu g_j^\tau$ for the vector and axial vector $U_A(1)_{L_\mu-L_\tau}$ are different. For comparison, the ordinary vector and our axial charge assignments are listed in Tab. \ref{tab:vector-axial-charges}. 
\begin{table}[htbp]
    \centering
    \begin{tabular}{c|cccc}
    \hline
     & $g_L^\mu$ & $g_R^\mu$ & $g_L^\tau$ & $g_R^\tau$ \\
    \hline
    vector & $g_X$ & $g_X$ & $-g_X$ & $-g_X$ \\
    axial & $g_X$ & $-g_X$ & $-g_X$ & $g_X$ \\
    \hline
    \end{tabular}
    \caption{Chiral coupling constants for left- and right-handed neutral-current interactions for vector and axial vector $U_A(1)_{L_\mu-L_\tau}$ models.}
    \label{tab:vector-axial-charges}
\end{table}
From this table, and combined with Eqs. \eqref{eq:mucollider-chiral-xsec} and \eqref{eq:mucollider-chiral-amplitude}, we can immediately see that a total-rate reinterpretation of the vector analysis does not capture the main axial effect. To capture this axial effect, we can take forward–backward asymmetry method, which is characterized as
\begin{equation}
A_{\rm FB}=
\frac{\sigma_{\ell\ell}(\cos\theta>0)-\sigma_{\ell\ell}(\cos\theta<0)}
{\sigma_{\ell\ell}(\cos\theta>0)+\sigma_{\ell\ell}(\cos\theta<0)}.
\label{eq:mucollider-afb}
\end{equation}
As an example, in Fig.~\ref{fig:muon-collider-angular}, this effect is illustrated. For $m_{A'}=1~{\rm TeV}$ and $g_X=0.05$, the vector benchmark changes the $|\cos\theta|<0.9$ total rate by about $5\%$, while the axial benchmark changes it by only about $2\%$. Conversely, the axial case shifts $A_{\rm FB}$ by $\Delta A_{\rm FB}\simeq-2.9\times10^{-2}$, larger than the vector shift in this benchmark. By implementing angular-bin or asymmetry analysis, the constraints on this axial $U_A(1)_{L_\mu-L_\tau}$ model can be derived for heavy $A'$, which makes the future collider channel complementary to the low-energy probes above. As outlined at the start of this section, robust exclusion limits require detailed detector-level Monte Carlo simulations. Such a treatment would largely replicate the analysis procedure of Ref. \cite{Huang:2021nkl}. For this reason, we omit a full collider simulation in present work and present a novel feasible search method instead.

\section{Conclusion}
\label{sec:conclusion}

In this work, we have constructed an anomaly-free and renormalizable axial $U_A(1)_{L_\mu-L_\tau}$ model. Since the charge assignment forbids the diagonal SM Yukawa operators, a singlet scalar and heavy vector-like leptons generate the charged-lepton masses through a universal-seesaw-like mechanism. After diagonalizing the mass matrices, we find that the heavy-lepton limit recovers the SM Higgs Yukawa couplings and leaves a dominantly axial light-lepton current, with only an $\mathcal O(\theta_R^2)$ vector admixture. The benchmark choice $y_{\mu\tau}=0$ is both compatible with charged-lepton-flavor-violation bounds and technically natural in the presence of an approximate family symmetry.
In the pure-axial limit, we derived and reinterpreted the leading constraints summarized in Fig.~\ref{fig:allconstraints}. For neutrino trident production, the full finite-$m_\mu$ calculation shows that the mediator shifts $C_A$, rather than $C_V$ as in the vector benchmark. Muon beam-dump production and $(g-2)_\mu$ are particularly sensitive to the longitudinal enhancement associated with axial-current nonconservation; the NA64$_\mu$ contour is therefore presented as a rate-level phenomenological recast, while $(g-2)_\mu$ acts as a consistency bound. Invisible pion and kaon decays probe the complementary regions below their respective kinematic endpoints, with the NA62 curve interpreted as an indicative acceptance-dependent recast. By contrast, for $m_{A'}\gg m_\mu$, axial and vector final-state-radiation rates converge, allowing the BaBar, Belle~II, and CMS four-muon limits to be reinterpreted through a controlled branching-ratio rescaling. Far below the mass range shown in Fig.~\ref{fig:allconstraints}, the coherent field sourced by solar neutrinos provides a complementary probe of ultralight $A'$ bosons through the day--night modulation of muon spin precession.

At a future muon collider, radiative return remains an efficient on-shell search, whereas the angular distribution of $\mu^+\mu^-\to\tau^+\tau^-$ provides a direct discriminator of the interaction structure. The axial contribution is more clearly exposed through angular bins or the forward--backward asymmetry than through the total rate, although a detector-level analysis is required to obtain a robust exclusion. In summary, an axial $L_\mu-L_\tau$ force cannot be obtained by simply replacing $g_V$ with $g_A$ in the conventional benchmark. Its low-energy behavior is tied to charged-lepton mass generation and axial-current nonconservation, while vector-to-axial rescaling is justified only for selected high-energy observables.

\section{Acknowledgments}
The work of J.G. is supported by the Postdoctoral Fellowship Program (Grade C) of China Postdoctoral Science Foundation under Grant No. GZC20252775.
The work of J.L. is supported by the National Science Foundation of China under Grant No. 12235001, No. 12475103
and State Key Laboratory of Nuclear Physics and Technology under Grant No. NPT2025ZX11. 
The work of X.P.W. is supported by National Science Foundation of China under Grant No. 12375095, and the Fundamental Research Funds for the Central Universities.
J.L. and X.P.W. thank the Asia Pacific Center for Theoretical Physics (APCTP), Pohang, Korea, for their hospitality during the focus program [APCTP-2025-F01], from which this work greatly benefited. J.L. and X.P.W. also thank the Mainz Institute for Theoretical Physics (MITP) of the PRISMA+ Cluster of Excellence (Project ID 390831469) for its hospitality and partial support during the completion of this work. We also acknowledge with appreciation the valuable discussions and insights provided by the members of the Collaboration of Precision Testing and New Physics.

\bibliographystyle{JHEP}
\bibliography{references}

\end{document}